\documentclass[doublecol,linenumbers]{epl2} 
\usepackage{amsmath}
\usepackage{cancel}
\usepackage{float}
\usepackage{ifsym}
\usepackage{color}
\usepackage{latexsym}
\usepackage{latexsym}
\usepackage{dcolumn}
\usepackage{epsfig}
\usepackage{amsfonts}
\usepackage{amsmath}
\usepackage{dsfont}
\usepackage{accents}
\usepackage[all,cmtip]{xy}
\usepackage{bm}
\usepackage{graphicx}
\usepackage{wasysym}
\usepackage{hyperref} 

\institute{                    
  \inst{1} First Institute - Address\\
  \inst{2} Second Institute - Address}

\title{Dynamic Mechanical Analysis of supercooled water in nanoporous confinement}
\author{Viktor~Soprunyuk\inst{1} \and Wilfried~Schranz\inst{1} \and Patrick~Huber\inst{2}}
\institute{\inst{1} University of Vienna, Faculty of Physics, Physics of Functional Materials,
Boltzmanngasse 5, A-1090 Wien, Austria\\
\inst{2} Institute of Materials Physics and Technology, Hamburg University of Technology (TUHH),
Ei\ss endorfer Str. 42, D-21073 Hamburg-Harburg, Germany}
\date{\today}

\pacs{64.70.P-}{Glass transitions of specific systems}
\pacs{65.40.De}{Thermal expansion; thermomechanical effects}
\pacs{81.05.Rm}{Porous materials}

\abstract{Dynamical mechanical analysis (DMA)(f=0.2 - 100 Hz) is used to study the dynamics of confined water in mesoporous Gelsil (2.6 nm and 5 nm pores) and Vycor (10 nm) in the temperature range from T=80~K to 300~K. Confining water into nanopores partly suppresses crystallization and allows us to perform measurements of supercooled water below 235~K, i.e. in water's so called "no man's land", in parts of the pores. Two distinct relaxation peaks are observed in tan$\delta$ around T$_1 \approx$ 145~K ($P_{\rm 1}$) and T$_2 \approx$ 205~K ($P_{\rm 2}$) for Gelsil 2.6 nm and Gelsil 5 nm at 0.2 Hz. Both peaks shift to higher T with increasing pore size $d$ and change with f in a systematic way, typical of an Arrhenius behaviour of the corresponding relaxation times. For $P_{\rm 1}$ we obtain an average activation energy of E$_a$=0.47 eV, in good agreement with literature values. 
It is suggested that $P_{\rm 1}$ corresponds to the glass transition of supercooled water far from pore walls, whereas $P_{\rm 2}$ reflects the dynamics of water molecules near the surface of the pores. The observation of a pronounced softening of the Young's modulus around 165~K (for Gelsil 2.6 nm at 0.2 Hz) is in agreement with a glass-to-liquid transition in the vicinity of $P_{\rm 1}$. 
In addition we find a clear-cut $1/{\rm d}$-dependence of the calculated glass transition temperatures which extrapolates to $T_{\rm g}$(1/d=0)=136~K, i.e. the traditional value of water.} 

\begin{document}
\maketitle

\section{Introduction}
Water is not only of fundamental importance for life, its behaviour is of great relevance for biology, geology, chemistry, physics and technology. Despite a long history of research many properties of water are still far from being understood \cite{Debenedetti2003}. Some of them, e.g. the location of its glass transition temperature T$_{\rm g}$ and the existence of a liquid-liquid phase transition \cite{Poole1992, Stanley2010} are rather controversially discussed in the literature \cite{Angell2004,Kohl2005,Angell2005,Hill2016, Cerveny2016}. Historically the glass transition temperature of amorphous solid water (ASW) and hyperquenched glassy water (HGW) was determined as $T_{\rm g} \approx 136 K$ \cite{Johari1987}, but since 2002, Angell, \emph{et al.} \cite{Angell2002,Angell2004} has raised question on this value and has proposed that $T_{\rm g}$ of water is between 165 and 180 K. Cerveny, \emph{et al.} \cite{Cerveny2004} proposed a glass transition at 160-165 K for bulk water and about 175 K for confined water. Oguni \emph{et al.} \cite{Oguni2011} suggested even a value of 210~K for T$_{\rm g}$ of bulk water.\\
Unfortunately it is impossible to follow the relaxation time of bulk water continuously down to T$_{\rm g}$, since it crystallizes not later than T$_{\rm H}$=235 K which corresponds to the homogeneous nucleation temperature. At high temperatures, i.e. above 235~K water is a very fragile liquid (Vogel-Fulcher relaxation time dependence) \cite{Ito1999}, while on the other side of the so called "no mans land" (150-235~K) it was found to be a "superstrong" liquid \cite{Novikov2013}. This difference in "fragility" for water was used by Ito, \emph{et al.} to propose the existence of a fragile-to-strong (FTS) transition in supercooled water near 228~K. During recent years, several authors have looked for a fragile-to-strong transition in confined water \cite{Bergman2000,Liu2005} and biomaterials \cite{Chen2006}. However, its interpretation is subject of intense debate \cite{Johari2000,Chen2006,Capaccioli2011}.\\
Confining water in mesoscopic environments is a way to suppress crystallization and even avoid it in pores smaller than about 2 nm in diameter \cite{Erko2011}. 
Here we present results of extensive Dynamic Mechanical Analysis (DMA) and Thermomechanical (TMA) measurements of water confined in mesoporous silica, Vycor and Gelsil with pore diameters $d$ of 10 nm (V10), 5 nm (G5) and 2.6 nm (G2) and discuss our results in the light of previous studies. 

\section{Experimental Results}
A diamond saw was used to cut samples of Vycor and Gelsil with typical sizes of 4$\times$2$\times$2 mm$^3$ for V10, 3$\times$1.5$\times$1.5 mm$^3$ for G5 and 2.5$\times$2$\times$2 mm$^3$ for G2. The samples were sanded to gain parallel surface plains. Cleaning was done in a 30\% H$_2$O$_2$ solution at 90$^o$C for 24~h, followed by drying at 120$^o$C in a high-vacuum chamber also for 24~h. Filling with distilled water was done by spontaneous imbibition \cite{Gruener2009}.\\
For thermal expansion measurements we used a TMA 7 (Perkin Elmer). To study the slow dynamics of confined supercooled water, we performed Dynamic Mechanical Analysis (DMA) measurements (Diamond DMA and DMA 7, Perkin Elmer) as a function of frequency $f$ (0.01--100 Hz) and temperature $T$ (80 -- 300 K). 

\begin{figure}
\centering
\includegraphics[width=6cm]{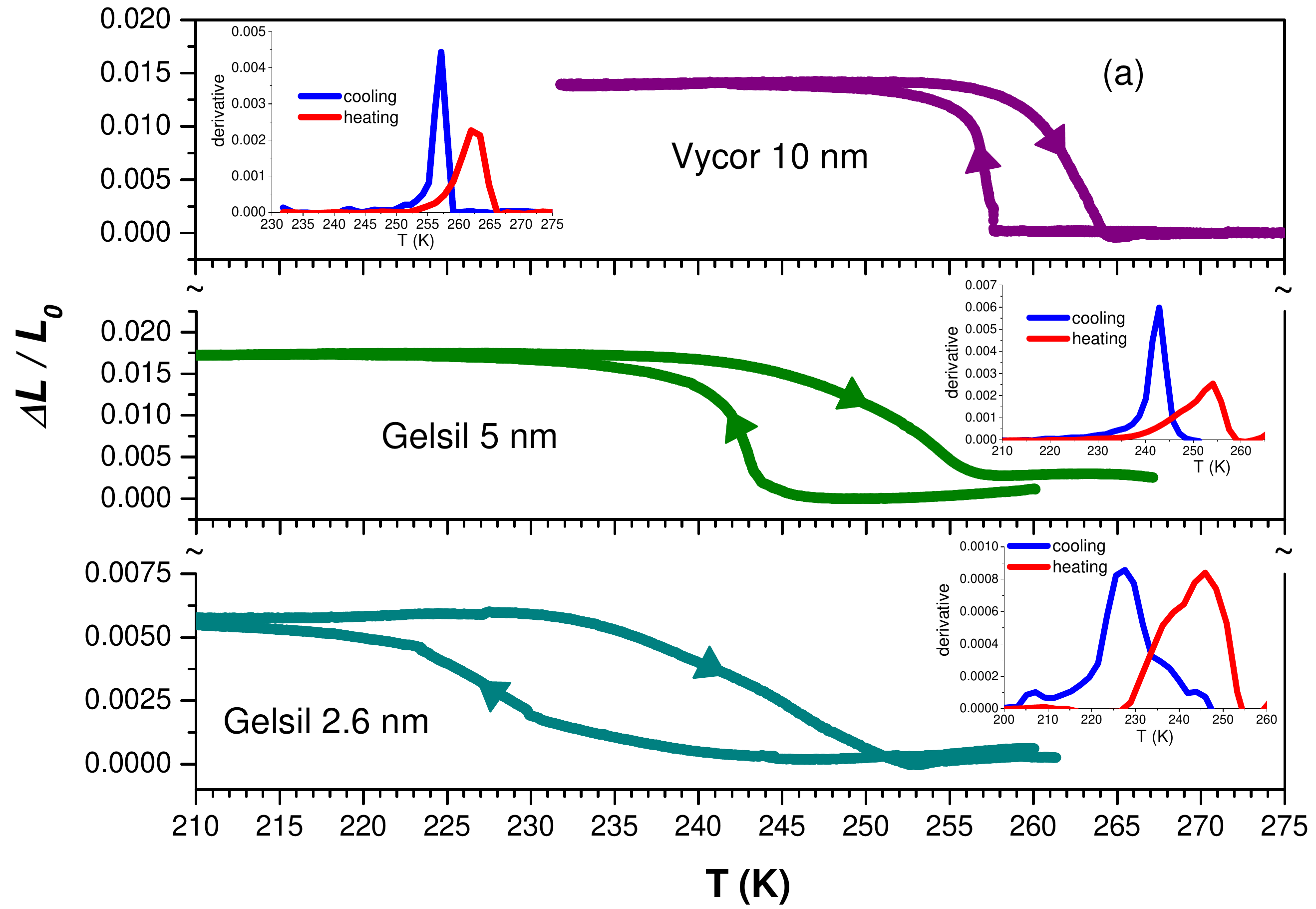}
\includegraphics[width=6cm]{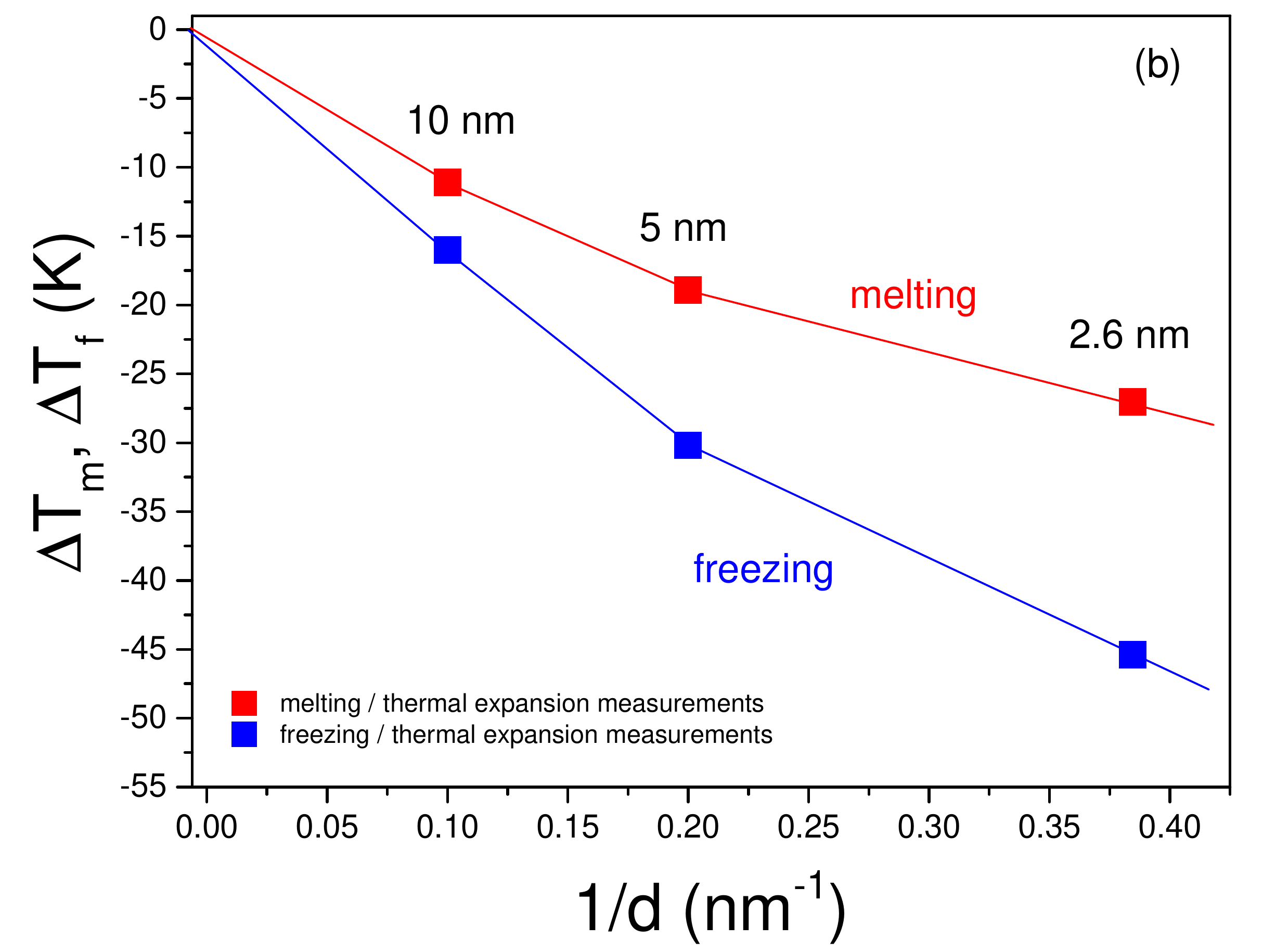}
\caption{(a) Thermal expansion of mesoporous silica with different pore sizes fully filled with water. (b) Confinement induced shifts of melting and freezing temperatures.}
\label{fig:thermal_expansion}
\end{figure}

The samples were mounted in a cell and rapidly cooled down. Two different cooling procedures, i.e. with a rate of about 1 K/min to 80 K as well as quenching the samples in liquid nitrogen has led to identical results in subsequent heating runs. The real $Y'$ and imaginary $Y''$ parts of the complex Young's modulus $Y^\ast=Y'+iY''$ were measured in parallel plate geometry. They are calculated from the elastic compliance tensor S$_{\rm ii}^{\ast}$, which is determined from the relation between the sample strain appearing in response to the applied dynamic force as well as the phase shift $\delta$ between dynamic force and sample strain. Details of the DMA method are given in Refs.\cite{Schranz1997,Salje2011}. 
Most of the measurements were performed by heating from 80~K with a heating rate of 1.5~K/min. To avoid breaking of the sample we stopped every heating run around 230~K, cooled down the sample to 80~K and started the next heating run with another measurement frequency. In this way we measured $Y^\ast$ as a function of $T$ and $f$.\\
We also measured thermal expansion of the samples during cooling and heating using a TMA7 (Perkin Elmer). The results are shown in Fig.~\ref{fig:thermal_expansion}. The corresponding melting and freezing temperatures shown in the inset of the Figure were calculated from the maxima of the derivatives of thermal expansion, i.e. of $\alpha =1/L_{\rm 0} \frac {\partial \Delta L}{\partial T}$.   

\begin{figure}
\centering
\includegraphics[width=7cm]{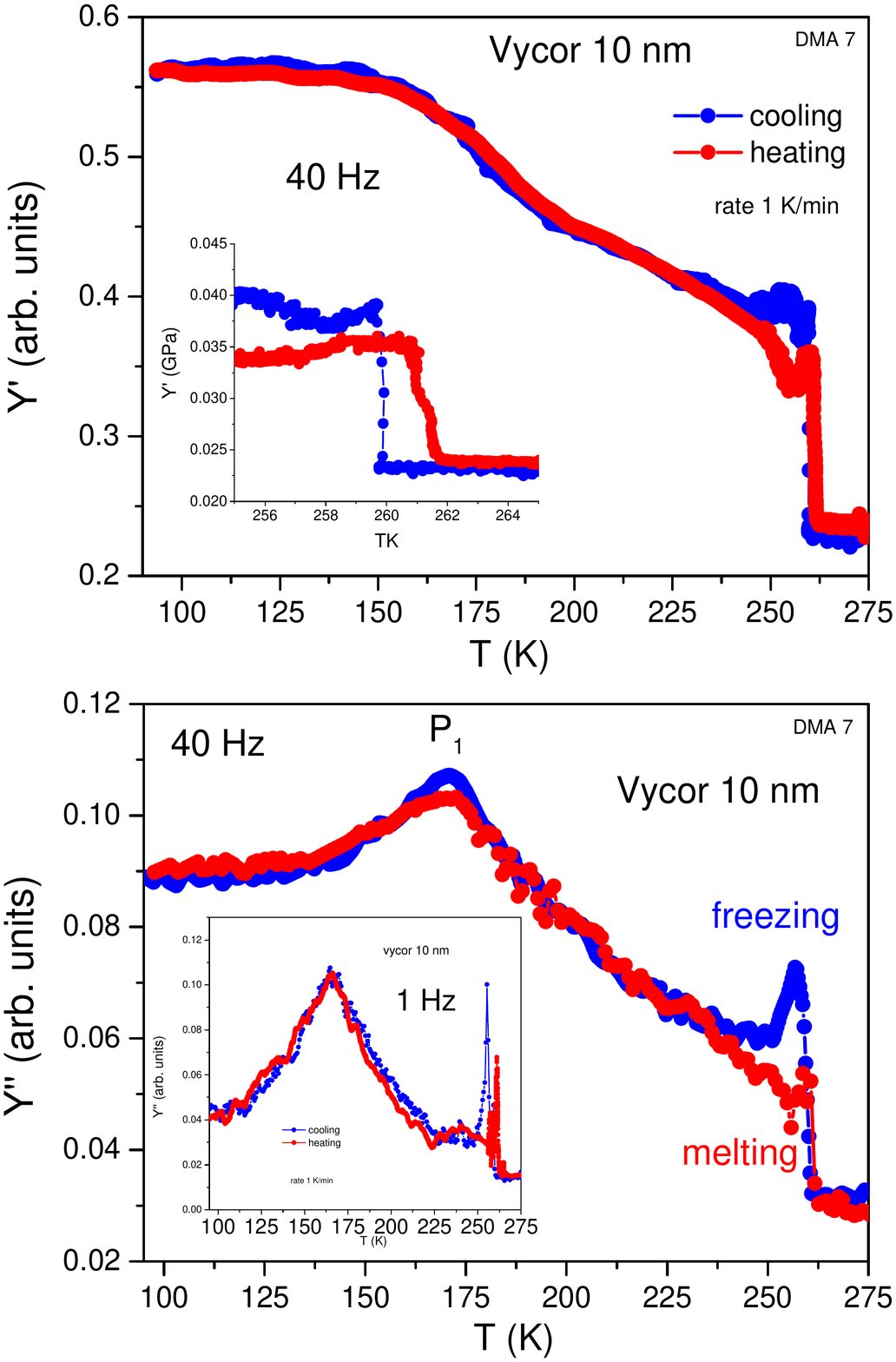}
\caption{Temperature dependence of real $Y'$ and imaginary $Y''$ parts of the complex Young's modulus $Y^{\ast}$ of V10 filled with water, measured at 
40 Hz with a DMA7 (Perkin Elmer). The inset in $Y'(T)$ displays the T-hysteresis. The inset below shows $Y''(T)$ measured at 1 Hz for comparison. Note that the heating-melting peak is independent of frequency, whereas $P_{\rm 1}$ shifts to lower T for lower f.}
\label{fig:Vycorfullrange}
\end{figure}

\begin{figure}
\centering
\includegraphics[width=6cm]{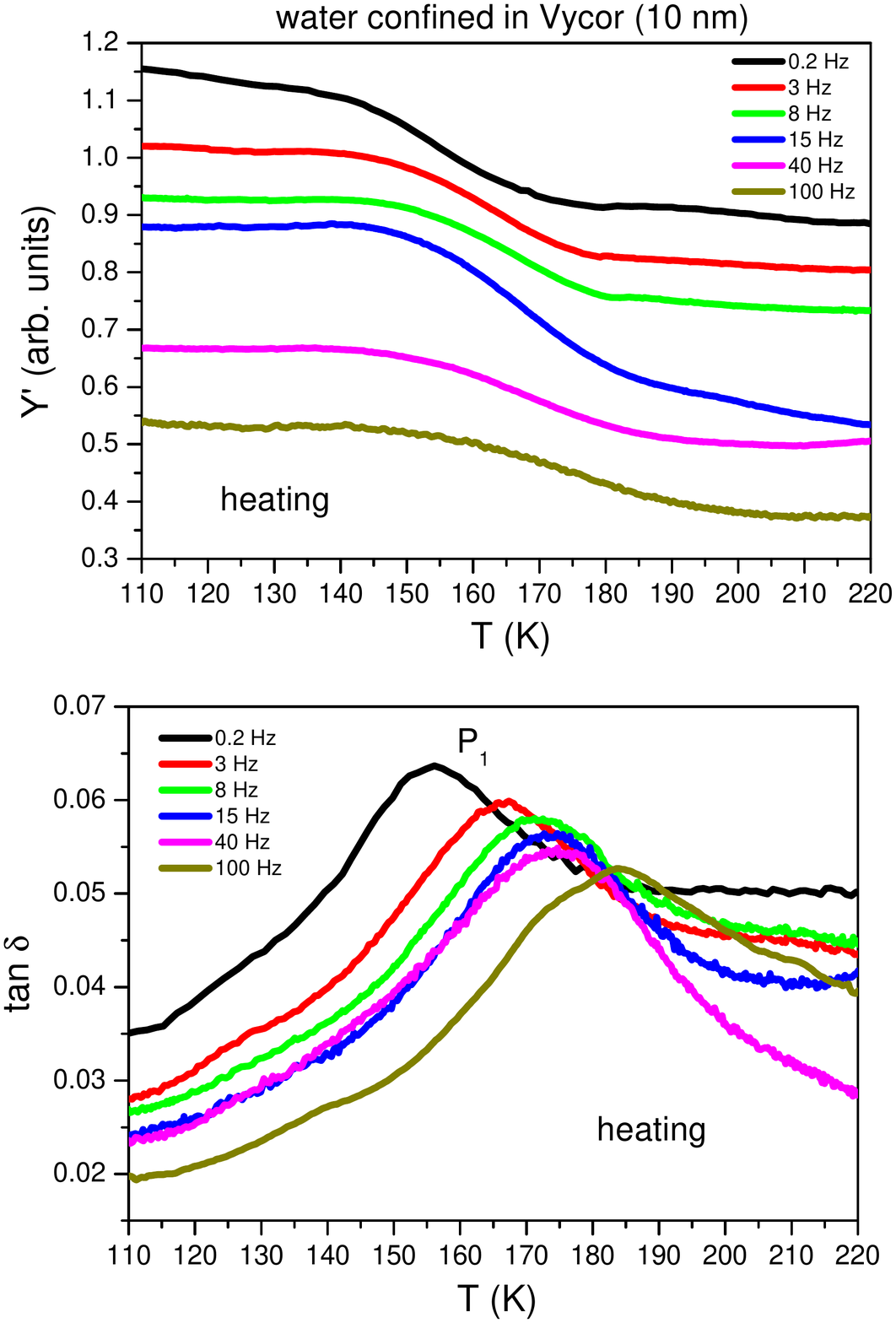}
\caption{Temperature dependence of Young's modulus $Y'$ and $\tan \delta$ of V10 filled with water, at various measurement frequencies. The curves are shifted for clarity.}
\label{fig:Vycorfrequency}
\end{figure}

One clearly observes a decrease of the freezing- and melting temperatures with decreasing $d$ in agreement with previous observations \cite{Findenegg2008} accompanied by a broadening of the transitions. From these data it is obvious that a considerable amount of water crystallizes even for the smallest $d$ of 2.6 nm.\\
Although these data are helpful to get a first clue on the state of the system, to learn more about the corresponding structural dynamics we have to inspect the results of DMA measurements. Fig.~\ref{fig:Vycorfullrange} shows a typical pattern of $Y'$ and $Y'' $ for water in V10 as a function of $T$ measured with a DMA7 during heating and cooling with a rate of 1~K/min at 40~Hz.  
The abrupt increase of $Y'$ around 265~K which is accompanied by a peak in $Y''$ relates to the freezing of water in the pores.  
At lower $T$ a second peak ($P_{\rm 1}$) at about 155~K (0.2 Hz) in $Y''$ is observed which is accompanied by an "S-shaped" anomaly in $Y'$, resembling a typical relaxation behaviour. To study the origin of $P_{\rm 1}$ in more detail we measured $Y'(T)$ and $\tan\delta (T)$ at various frequencies using a Diamond DMA (Perkin Elmer). Fig.~\ref{fig:Vycorfrequency} shows the results for water in V10. One observes a clear shift of $P_{\rm 1}$ to higher temperatures with increasing $f$ in contrast to the melting peak (Fig.~\ref{fig:Vycorfullrange}) which turns out to be independent of frequency. 

\begin{figure}
\centering
\includegraphics[width=7cm]{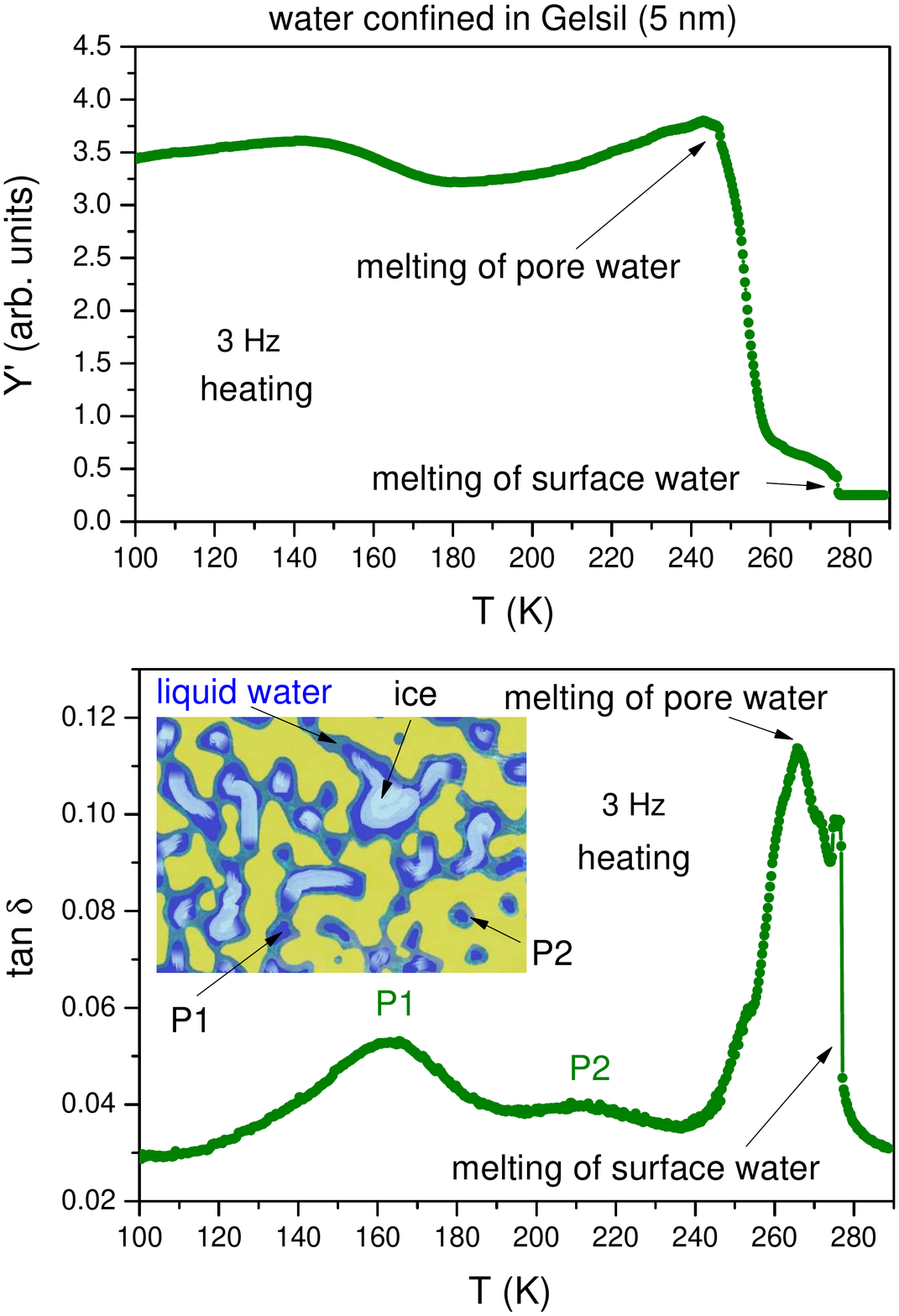}
\caption{Temperature dependence the Young's modulus $Y'$ and $\tan \delta$ of G5 filled with water, measured at 3~Hz. The inset shows a sketch of the "pea-in-pod" model of ice formation proposed in \cite{Sellevold1976}.}
\label{fig:Gelsil5fullrange}
\end{figure}

Fig.~\ref{fig:Gelsil5fullrange} displays the $T$-dependence of $Y'$ and $\tan \delta$ for G5 filled with water, measured at 3~Hz. In addition to the  melting of water in the pores one observes a second process at 273~K which corresponds to the melting of surface water. 
Similar as for water in V10 a clear relaxation peak ($P_{\rm 1}$) around 165~K (3 Hz) is observed, which is accompanied by a distinct minimum in $Y'$ around 180~K. An additional peak $P_{\rm 2}$ at about 215~K is also found for G2. Its possible origin will be discussed later.  

\begin{figure}
\centering
\includegraphics[width=6cm]{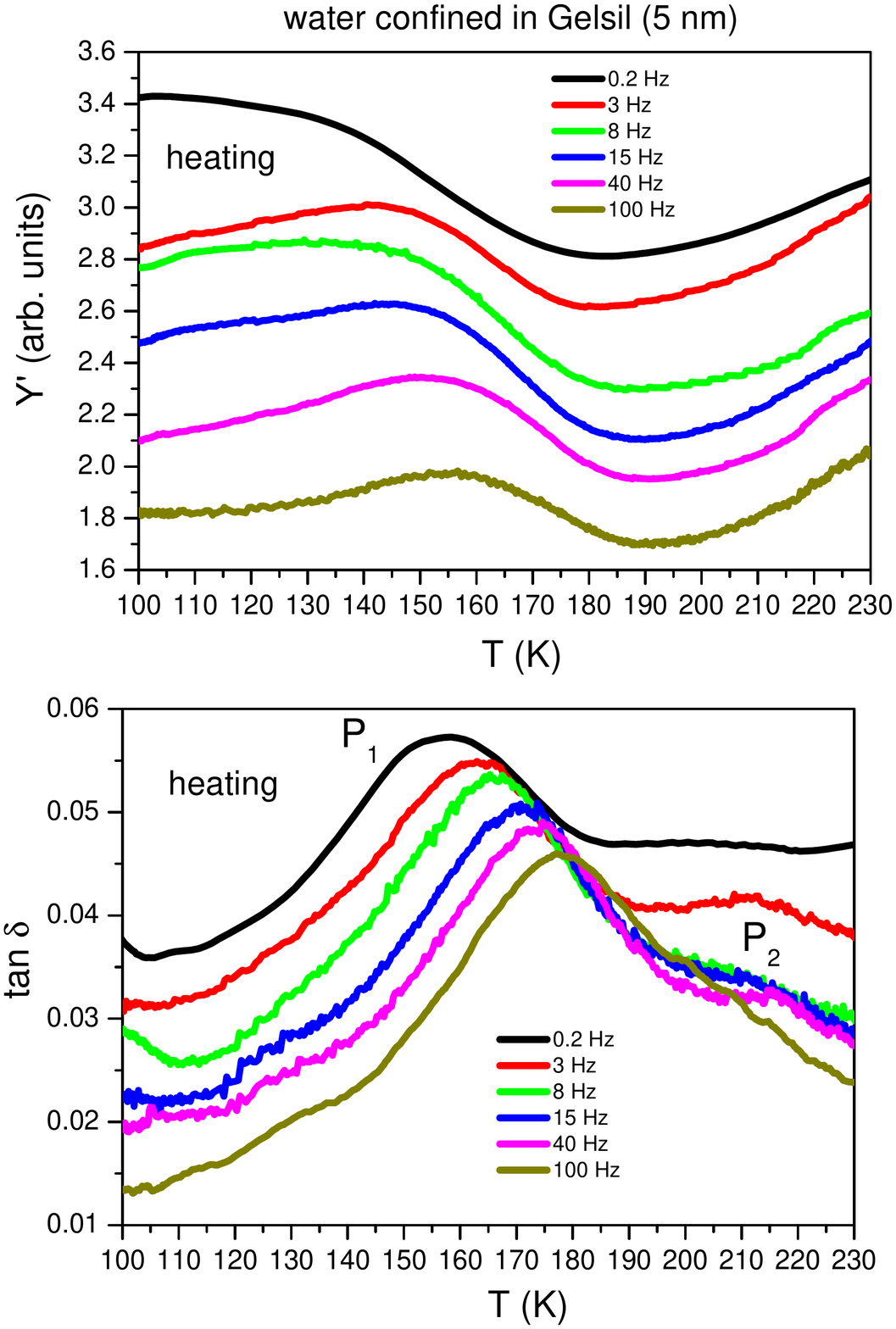}
\caption{Temperature dependence of the Young's modulus Y' and tan$\delta$ of G5 filled with water, measured at various frequencies. The curves are shifted for sake of clarity.}
\label{fig:Gelsil5frequency}
\end{figure}

\begin{figure}
\centering
\includegraphics[width=6cm]{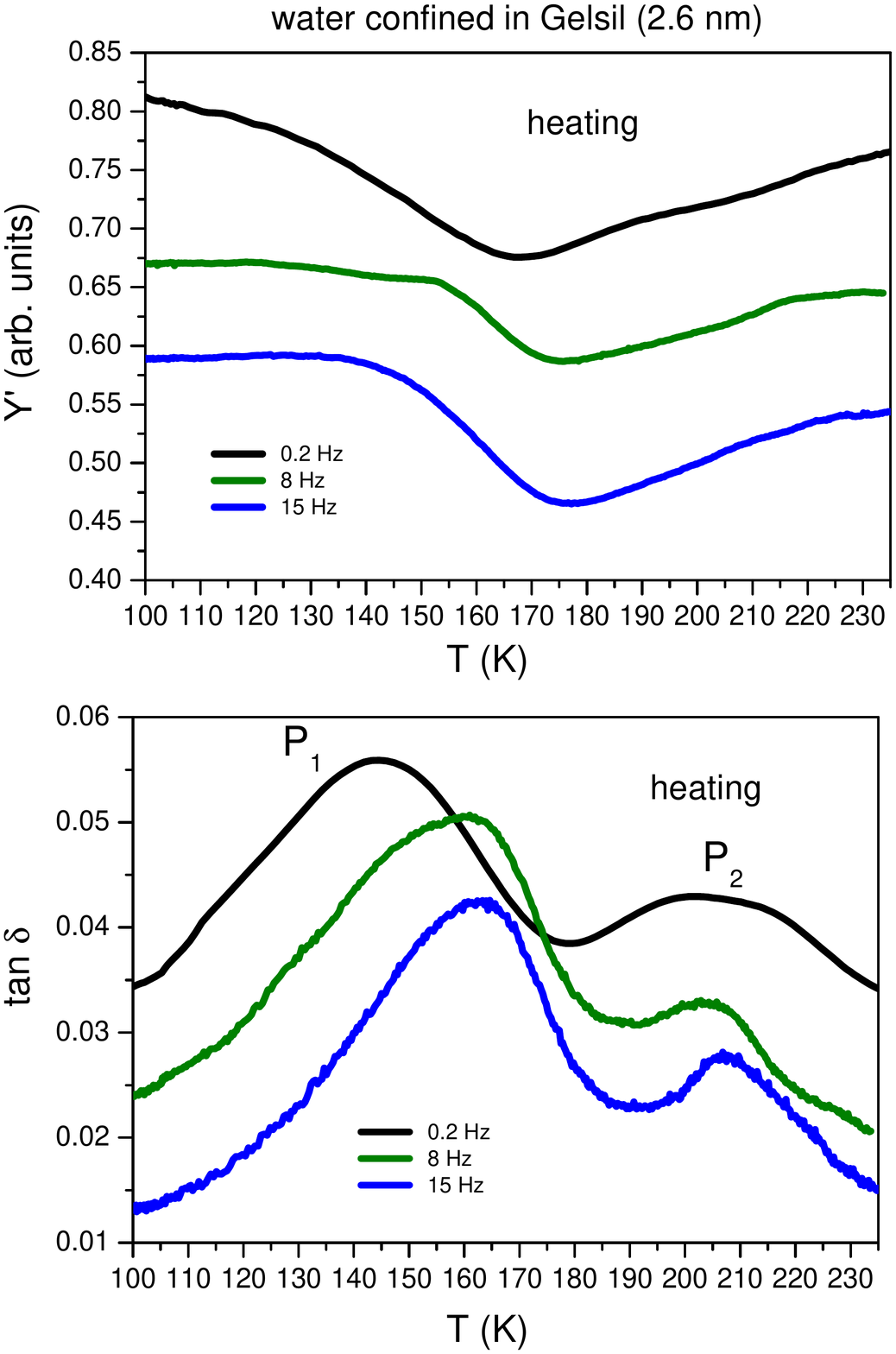}
\caption{Temperature dependence of the Young's modulus Y' and tan$\delta$ of G2 filled with water, measured at various frequencies.}
\label{fig:Gelsil2frequency}
\end{figure}

Fig.~\ref{fig:Gelsil5frequency} shows the $f$-dependence of $Y^*$ of G5 filled with water. The overall behaviour is very similar to water in V10, with the addition of a second peak $P_{\rm 2}$ around 215~K (3 Hz) and a very pronounced minimum in $Y'$ around 180~K (at 3 Hz), which also depends on $f$.\\    
In G2 filled with water the additional peak $P_{\rm 2}$, which appears now at lower temperature ($\sim$200~K) as compared to G5, is rather pronounced, see Fig.~\ref{fig:Gelsil2frequency}. In addition the minimum in $Y'$ around 170~K (at 3~Hz) is now also strongly developed.    

\section{Discussion}
\begin{figure}
\centering
\includegraphics[width=6cm]{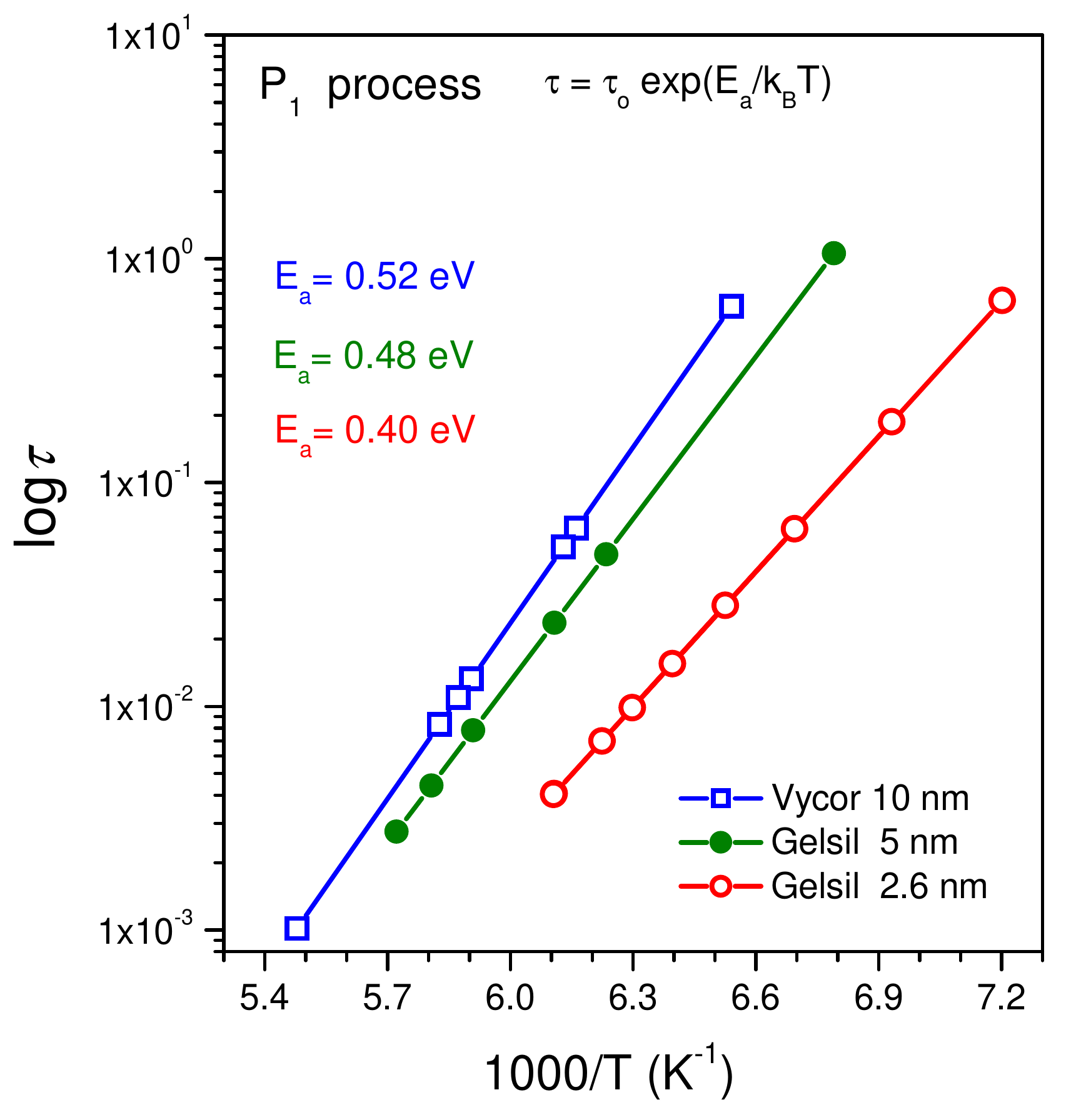}
\caption{Arrhenius plots for water confined in V10 and Geslis 5 nm and 2.6 nm.}
\label{fig:Arrhenius}
\end{figure}

\begin{figure}
\centering
\includegraphics[width=5cm]{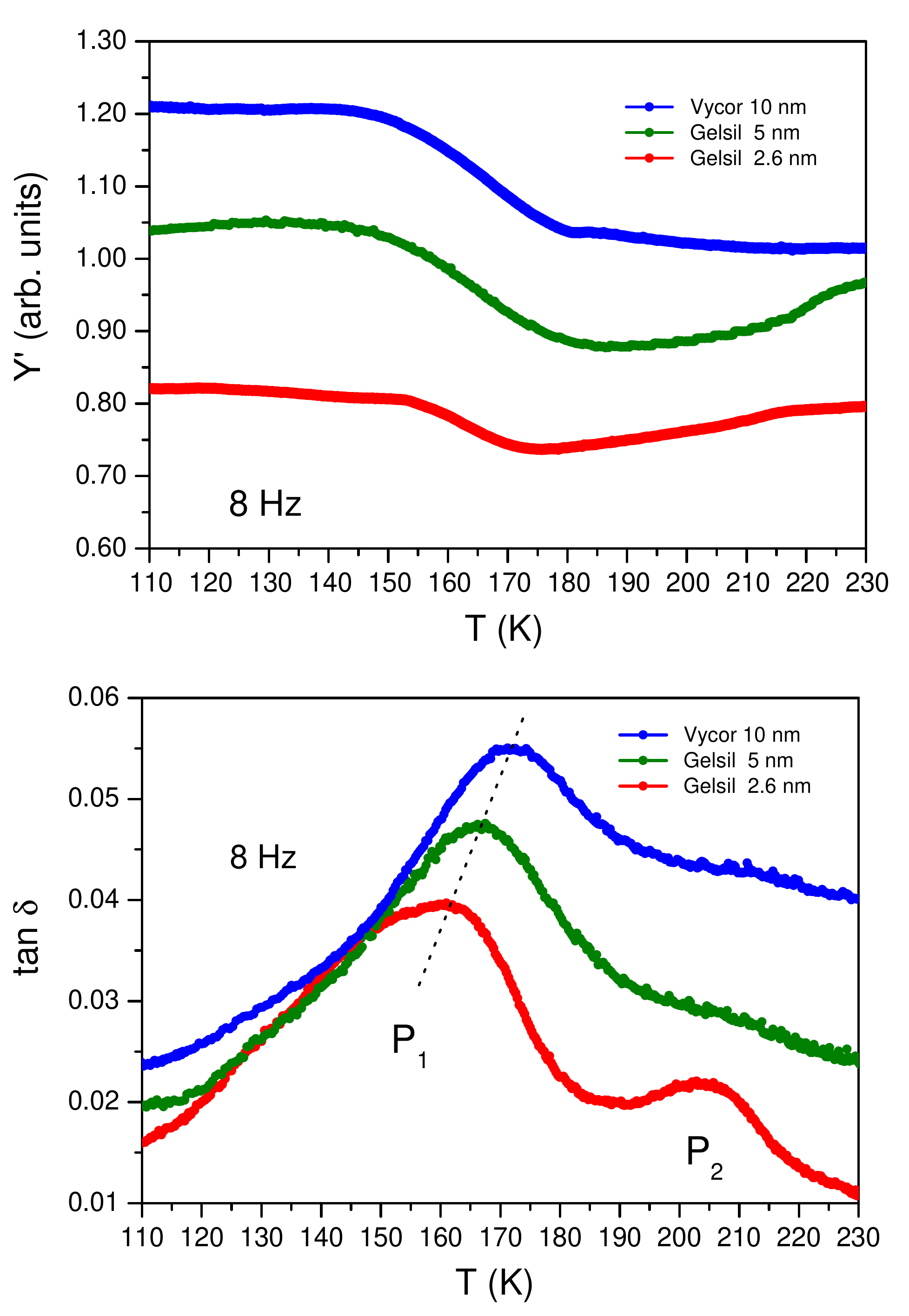}
\caption{Comparison of the temperature dependencies of Young's modulus and tan$\delta$ of V10, G5 and G2 filled with water.}
\label{fig:peakshifts}
\end{figure}

To analyse the data, especially the origin of the $P_{\rm 1}$ process, we determined time constants $\tau=1/(2\pi\nu_{\rm m})$ from the loss peak frequencies $\nu_{\rm m}$ and summarize the results in an Arrhenius diagram (Fig.~\ref{fig:Arrhenius}).
The straight lines in Fig.~\ref{fig:Arrhenius} indicate that the relaxation times corresponding to the $P_{\rm 1}$-process follow thermal activation according to 
$\tau=\tau_{\rm 0} exp(\Delta E/k_{\rm B}T)$ with $d$-dependent activation energies.  
This results in a downshift of the peaks $P_{\rm 1}$ and $P_{\rm 2}$ with decreasing $d$ (Fig.\ref{fig:peakshifts}). 
The present activation energies (Fig.\ref{fig:Arrhenius}) compare well with literature data \cite{Pissis1994,Swenson2010,Sjostrom2008}. 
Sjostrom, \textit{et al.} \cite{Sjostrom2008} performed calorimetry and dielectric measurements of water confined in MCM-41 with $d=$2.1 (C10) and 3.6~nm (C18). They found that in C10 no ice formation occurs, whereas for C18 ice formation becomes substantial. However, not all water crystallizes in C18. Some regions of capillary condensed water remain liquid even at low $T$. The authors attributed the dielectrically observed process with activation energy of $E_{\rm a} \approx$ 0.47 eV for C10 and C18 to the relaxation of amorphous water.\\
Cerveny, \textit{et al}. \cite{Cerveny2004} performed broadband dielectric spectroscopy measurements of supercooled water confined in clay ($d=$1.5~nm) and in white bread and compared it to various biological systems. They also obtained an average activation energy of $E_{\rm a} = (0.46 \pm 0.04) eV$ and the temperature where $\tau=100$ s was extrapolated to $T$(100s)=(139 $\pm$ 3)K. Traditionally this temperature was associated with the glass-to-liquid transition \cite{Johari1987,Bergman2000,Cerveny2016} of supercooled water, but recently this was doubted.

Cerveny, \textit{et al}. \cite{Cerveny2004} related the $P_{\rm 1}$ - relaxation to a local process of the Johari - Goldstein \cite{Goldstein1970} type. They located the glass transition at $T_{\rm g}$ = 160 - 165~K for bulk water and about 175~K for confined water (depending on the confining system). 
We do not think that the process $P_{\rm 1}$ which we detect in our DMA measurements is a local process. The pronounced minimum in $Y'$ around 170~K for 0.2 Hz (Fig.\ref{fig:Gelsil2frequency}) in G2, which in G5 occurs at a higher $T$ of about 180~K (Fig.\ref{fig:Gelsil5frequency}), suggests that a considerable amount of water in the pores is in a liquid (probably ultraviscous) state which below this mimimum in $Y'$ transforms to glass. 
Support for this scenario comes from recent neutron scattering data on amorphous solid water \cite{Hill2016}, which revealed the onset of long range diffusive motion of water molecules at T$>$ 121~K, marking the onset of a glass transition with its endpoint at $T_{\rm g}$=136~K. 
In our setup this transformation to glass is accompanied by an increase of the Young's modulus. Unfortunately, at present we cannot calculate real numbers for the hardening-or softening effects caused by the glass-to-liquid transition, because we do not know the actual fraction of crystallization for the different pore sizes. However, for a semi-quantitative analysis we adopt the so called "pea-in-pod" model of ice formation in Vycor, proposed earlier by Sellevold and Radjy \cite{Sellevold1976}. In their model they assume that inside the complex pore channels of Vycor (in their case $d=4$~nm) the thicker regions contain crystalline ice, whereas in the thinner channels supercooled water can still exist. A sketch of this situation is given here in the inset of Fig.\ref{fig:Gelsil5fullrange}. 
The volume fraction of crystal vs. liquid water can vary with pore size. I.e. in Vycor 4 nm about 55\% of water was found to consist of ice \cite{Antoniou1964,Taschin2016}. With lowering temperature the water transforms to glass and cements the ice and the silica matrix effectively together, thereby increasing the Young's modulus (Figs \ref{fig:Gelsil5frequency} and \ref{fig:Gelsil2frequency}).

\begin{figure}
\centering
\includegraphics[width=8.7cm]{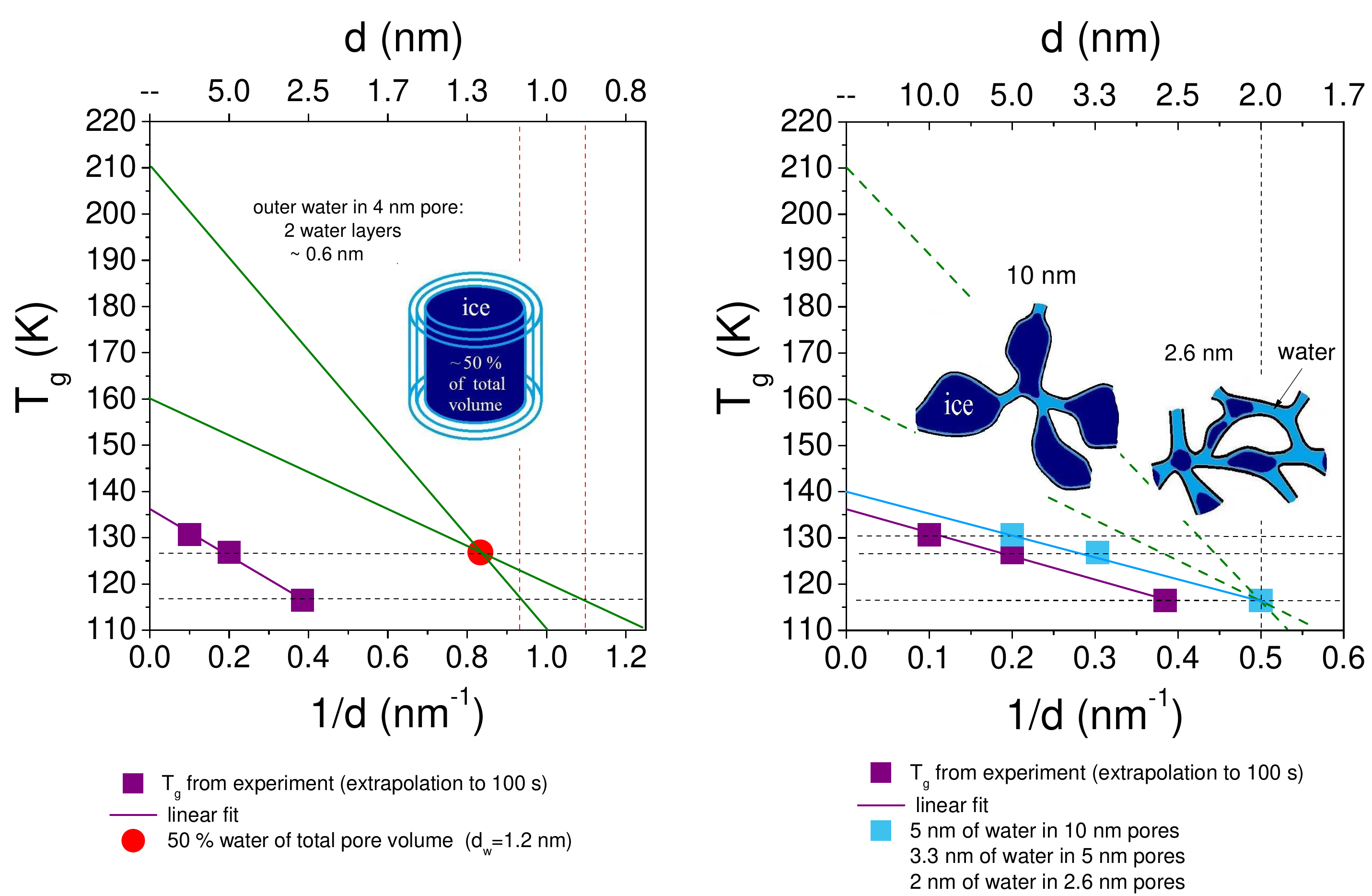}
\caption{Purple squares represent the measured size dependence of glass transition temperatures $T_{\rm g(d)}$ of supercooled water in Vycor and Gelsils and an extrapolation to T$_{\rm g}^{\rm bulk}=136~K$. For the two pictures two different estimates for effective amounts of supercooled water in various pore sizes are shown. The left picture corresponds to a "core-shell" model of ice-water in pores as e.g. used in \cite{Taschin2016}. In the right picture the relative volume of supercooled water is calculated from Fig.\ref{fig:thermal_expansion} and a "pea-in-pod" model \cite{Sellevold1976} is used to estimate $d_{\rm eff}$.}
\label{fig:ddependence}
\end{figure}

As shown in Fig.\ref{fig:peakshifts} the glass-to-liquid transition depends on the pore size. Extrapolating the relaxation times obtained for water in different confinements (Fig.\ref{fig:Arrhenius}) to 100~s yields a perfect $1/d$-dependence of $T_{\rm g(d)}$. A naive extrapolation to $1/d \rightarrow 0$ 
intercepts at $T_{\rm g}^{\rm bulk}$ = 136~K, in agreement with the traditional glass transition temperature of supercooled bulk water. However, since a considerable amount of water in the pores is crystalline \cite{Sellevold1976,Antoniou1964,Taschin2016}, such a simple extrapolation is questionable.\\ 
Taking into account, that the effective average available volume of supercooled water is smaller than the nominal one, we are dealing with an effective diameter $d_{\rm eff}$. As long as we do not know the exact amount of supercooled water and its geometrical location for different $d$, we cannot calculate the corresponding $d_{\rm eff}$, implying that we are not in the position to extrapolate our T$_{\rm g}$'s to obtain the bulk glass transition temperature of supercooled water. Let's discuss several scenarios in the light of complementary literature data. Fig.\ref{fig:ddependence} shows a comparison of two quite different situations. In both figures we plotted the glass transition temperatures (purple full squares) determined from the measured peak shifts extrapolated to 100~s. The purple line is a linear fit which extrapolates to $T_{\rm g}^{\rm bulk}$=136~K. Now the problem appears how to estimate the real effective size of supercooled water in the pores. One possibility would be to use a "core-shell" model of ice and water, which for about 50\% of ice leads to a value of 1.2~nm for the effective thickness of supercooled water \cite{Taschin2016} for Vycor 4 nm. In this way we should renormalize the effective size from 5 nm to 1.2 nm. The effect is shown in the left picture of Fig.\ref{fig:ddependence}. We see that even if we assume a very high glass transition temperature of 210~K \cite{Oguni2011}, we would get renormalized $d_{\rm eff}$-values for water in G2 which are hardly acceptable. We therefore believe, it is more reasonable to assume that at least in such disordered pore systems like Vycor and Gelsil, the ice and the supercooled water is not simply arranged in core and shell, but in a more complicated way, as e.g. was assumed in the "pea-pod"-model of ice formation \cite{Sellevold1976}. Such a partitioning is shown in the right picture of Fig.\ref{fig:ddependence}. In this case the renormalization of $d$ to $d \rightarrow d_{\rm eff}$ turns out to be much smaller. In fact, we can use our thermal expansion data (Fig.\ref{fig:thermal_expansion}) to get a rough estimate of at least the volume ratios of ice in the different pores, although we cannot estimate its location. Using the different values for $\Delta L_{\rm 0}/L_{\rm 0}$ in Fig.\ref{fig:thermal_expansion} we calculate the relative volume changes due to ice formation in the various pores. Together with the porosities of Gelsils and Vycor \cite{Koppensteiner2010}, we obtain $\Delta V/V_{\rm 0}$=8\% for V10, 6.7\% for G5 and 4.5\% for G2. Since the volume change of bulk water at freezing is about 9\% a simple calculation yields for the relative volume $V_{\rm sw}/V_{\rm pore}$ of supercooled water in the different pores $\approx$11\% for V10, $\approx$ 30\% for G5 and $\approx$50\% for G2. Making the assumption that $d_{\rm eff}$=$(V_{\rm sw}/V_{\rm pore})^{\rm 1/3}\times d$, we obtain for the renormalized $d_{\rm eff} \approx$ 5~nm for V10, 3.3 nm for G5 and 2 nm for G2. An extrapolation (blue line in Fig.\ref{fig:ddependence}) would then lead to a bulk $T_{\rm g}$ that is not far from the classical value of 136~K. From these considerations one realizes already that it is impossible to pin down a value of the bulk glass transition temperature of water from values that are determined from confined water without knowing where the supercooled water is located. However, it seems that the present data are in favour of the generally accepted value of $T_{\rm g} \approx$ 136 K for bulk water.\\ 
Finally, let's say a few words about the possible origin of the observed $P_{\rm 2}$-process (Figs.\ref{fig:Gelsil5fullrange},\ref{fig:Gelsil5frequency},\ref{fig:Gelsil2frequency}). A possible explanation can be found by a comparison of the present data to previous DMA measurements of molecular glass forming liquids, e.g. salol \cite{Koppensteiner2008,Koppensteiner2010}, toluene and o-terphenyl \cite{Schranz2010} in confinement. For salol confined in Vycor \cite{Schranz2007} and Gelsil \cite{Koppensteiner2008} we obtained results which resemble the present behaviour of water in very detail especially in Gelsils. At low temperatures we found two relaxation peaks at $T_{\rm g1} < T_{\rm g2}$ in $Y''$ accompanied by a double "S-shaped" temperature dependence of $Y'$. In these systems, we could identify the two peaks unambiguously. The one at $T_{\rm g1}$ was assigned to a glass transition of molecules in the core of the pores, whereas the other one at $T_{\rm g1}$ originates from molecules close to the pore walls. Due to the strong attractive interaction of salol-molecules with the pore wall the dynamics of the interfacial molecules is considerably slowed down, resulting in an increased $T_{\rm g}$. We think that a similar mechanism is also responsible for the $P_{\rm 2}$-relaxation process observed here for water in G2 and G5. 
Although, in the present case the core of the pores consists of ice, there are regions connecting the pores, where besides the water near pore walls exhibiting slowed down dynamics there is a substantial amount of supercooled water that is sufficiently away from the pore walls to exhibit faster dynamics ($P_{\rm 1}$-process). 
Since $P_{\rm 2}$ originates from an interface effect, it is perspicuous that it is not so much pronounced for V10. This is corroborated by recent computer simulations \cite{Klameth2013} and experiments \cite{Kityk2014}, where a strong slowing down of the dynamics of water and glass-forming methanol molecules were found, when approaching hydrophilic pore walls.
Note, however, that $P_{\rm 2}$ occurs close to the temperature anticipated for the liquid-liquid transition \cite{Poole1992, Cerveny2016} between a high- and low-density liquid phase in bulk water. Thus, we can not exclude that $P_{\rm 2}$ is a signature of this transition occurring here for interfacial water. This interpretation would be in agreement with conclusions in a previous calorimetric and neutron scattering study on water in Vycor \cite{Zanotti2005}.\\
Summarizing, we have shown that Dynamic Mechanical Analysis technique provides useful complementary information on the low frequency dynamics of water in nanopores. Together with other techniques providing the amount and location of supercooled water in nanoporous confinement we may in future be able to extrapolate these properties to bulk water.

\acknowledgments We acknowledge financial support from the Austrian Science Fund (FWF) Grant No. P28672-N36.

\end{document}